# Light focusing and additive manufacturing through highly scattering media using upconversion nanoparticles

Qianyi Zhang[1], Antoine Boniface[1], Virendra K. Parashar[2], Martin A. M. Gijs[2], and Christophe Moser[1, *]

[1] Laboratory of Applied Photonics Devices, School of Engineering, Institute of Electrical and Micro Engineering, Ecole Polytechnique Fédérale de Lausanne, Lausanne, Switzerland

[2] Laboratory of Microsystems LMIS2, School of Engineering, Institute of Electrical and Micro Engineering, Ecole Polytechnique Fédérale de Lausanne, Lausanne, Switzerland

[*] To whom correspondence should be addressed. e-mail: christophe.moser@epfl.ch

# Abstract

Light-based additive manufacturing holds great potential in the field of bioprinting due to its exceptional spatial resolution, enabling the reconstruction of intricate tissue structures. However, printing through biological tissues is severely limited due to the strong optical scattering within the tissues. The propagation of light is scrambled to form random speckle patterns, making it impossible to print features at the diffraction-limited size with conventional printing approaches. The poor tissue penetration depth of ultra-violet or blue light, which is commonly used to trigger photopolymerization, further limits the fabrication of high cell-density tissue constructs. Recently, several strategies based on wavefront shaping have been developed to manipulate the light and refocus it inside scattering media to a diffraction-limited spot. In this study, we present a high-resolution additive manufacturing technique using upconversion nanoparticles and a wavefront shaping method that does not require measurement from an invasive detector, i.e., it is a non-invasive technique. Upconversion nanoparticles convert near-infrared light to ultraviolet and visible light. The ultraviolet light serves as a light source for photopolymerization and the visible light as a guide star for digital light shaping. The incident light pattern is manipulated using the feedback information of the guide star to focus light through the tissue. In this way, we experimentally demonstrate that near-infrared light can be non-invasively focused through a strongly scattering medium. By exploiting the optical memory effect, we further demonstrate micro-meter resolution additive manufacturing through highly scattering media such as a 300-μm-thick chicken breast. This study provides a proof of concept of high-resolution additive manufacturing through turbid media with potential application in tissue engineering.

# Introduction

Bioprinting, a cutting-edge technology that merges biology and additive manufacturing, has revolutionized the field of tissue engineering[1,2]. This innovative approach allows for the precise deposition of biomaterials, cells, and growth factors to fabricate complex, functional tissues and organs[3–5]. As a result, bioprinting has opened new frontiers in tissue engineering, offering potential solutions for broad applications including disease modeling[6,7], drug testing[8,9], and regenerative medicine[1,10]. Common bioprinting methods include inkjet printing[11], extrusion-based printing[12], laser-induced forward transfer[13], and light-based additive manufacturing[14–16]. Bioprinted implants typically involve a surgical intervention for the implantation[17,18] or for direct in-situ biofabrication at the exposed site[19,20], which poses inherent challenges and risks.

To address these limitations, non-invasive and minimally invasive bioprinting has emerged as a powerful solution by offering the possibility of creating functional biological constructs bypassing invasive surgical procedures[21–24]. Light-based additive manufacturing, which employs light to solidify resins without the need for direct material deposition, is particularly well positioned compared to other minimally invasive bioprinting techniques thanks to light and its possibility of delivering energy through tissues. More precisely, the light energy is sent through the tissue to initiate photopolymerization of the injected bio-ink and transform it into desired structures. Light transport in biological tissues is determined by their absorption and scattering properties. Ultra-violet (UV) or blue light, which is commonly used in photopolymerization, shows poor tissue penetration depth and is not favorable for non-invasive bioprinting. The near-infrared (NIR) window with a wavelength ranging from 650 to 1350 nm, offers deeper penetration into biological tissues with less significant attenuation because of its longer wavelength (less scattering) and the lack of absorption from biological molecules[25]. Therefore, NIR light is well-suited for *in vivo* imaging[26–29] and therapeutic applications[29–31] that require light to reach target areas deep within the body. It can induce photopolymerization via two-photon absorption[32] or upconverting process[33] and has already been demonstrated in non-invasive additive manufacturing[21–23].

Although NIR light is transmitted more efficiently through tissues, scattering still scrambles the propagating light field to form complex speckle patterns, preventing focusing the light to a tiny spot for a well-confined delivery of light energy. This greatly impacts the resolution (from 1 μm in the absence of tissue to tens or hundreds μm depending on tissue properties and thickness) and the fidelity of non-invasive printing[21,23]. In the field of optical imaging, several strategies based on wavefront shaping have been developed to manipulate the light and refocus it through scattering media to a diffraction-limited spot[34–40]. These techniques utilize feedback signals obtained behind the scattering media to spatially modulate the input light in phase and amplitude. For non-invasive light focusing, fluorescent or acoustics signals emanating inside or behind the scattering medium can be measured from the same side as the light delivery[37,38,40]. However, these techniques only provide the wavefront information of one target location at a time, and it is time-consuming to refocus at each voxel to be printed. Fortunately, the scattered optical field preserves a certain degree of correlation, which is commonly referred to as the optical memory effect[41,42]. When an input wavefront reaching a scattering medium is shifted (or tilted) within a certain distance (or angle), the output wavefront propagating through the medium is equally shifted (or tilted). In thick biological media, where scattering is anisotropic (anisotropic factor g usually ranges from 0.9 to 0.98[25]), the range of tilt/tilt memory effect becomes minimal (50-μm-thick tissues around 3-8 mrad[43]) but strong shift/shift correlations are still observed[42]. In this way, the focal spot can be shifted through the scattering medium before it becomes too dim so that the next focusing optimization can be generated in a time-efficient manner[44]. The scattering effect of the tissue is thus corrected during the printing using sparse focusing, which significantly speeds up the printing as compared with optimizing at every subsequent spot.

In this study, we develop a micro-meter resolution additive manufacturing technique through a highly scattering medium assisted by upconversion nanoparticles (UCNPs). As the UCNP generates fluorescence of different wavelengths under the illumination of NIR light, it acts not only as a secondary UV source for photopolymerization but can also be used as a guide star for the feedback loop to refocus light through the scattering media. Then, the focal spot is scanned through the scattering medium using sparse focusing. Based on this technique, we are able to print high-resolution (2 μm) structures through a holographic diffuser and a chicken tissue of thickness 300 μm. These results demonstrate high-resolution additive manufacturing through strongly scattering media and suggest potential applications in non-invasive biomedicine.

## Results

We designed our non-invasive additive manufacturing system based on wavefront shaping, as illustrated in Fig. 1. A NIR beam at 976 nm is first modulated in amplitude by a digital micro-mirror device (DMD) and directed through a scattering medium (holographic diffuser or chicken tissue) into the resin. The resin contains hydrogel monomers of gelatin methacryloyl

(gelMA) and UCNPs coated with the UV light photoinitiator lithium phenyl-2,4,6-trimethylbenzoyl-phosphinate (LAP). UCNPs emit UV and visible fluorescence under the illumination of 976-nm light. The visible fluorescence (440 nm < λ < 550 nm), which is not absorbed by LAP, is back-scattered by the scattering medium and epi-detected by a single-photon avalanche diode (SPAD), providing the feedback signal for the optimization of the spatial light modulation (binary DMD pattern). Then, the optimized DMD pattern is displayed and refocuses the NIR light through the scattering medium to a diffraction-limited spot within the resin, which is shifted together with the scattering medium to induce photopolymerization along the designed path. After a lateral shift of the sample determined by the size of the memory effect (here around 5 μm), the DMD pattern is re-optimized to focus light again and this scanning process is repeated until the printed part is complete. The resulting spatial distribution of the NIR light at the focal plane is inspected by a camera placed on the distal side of the sample, for observation purposes only.

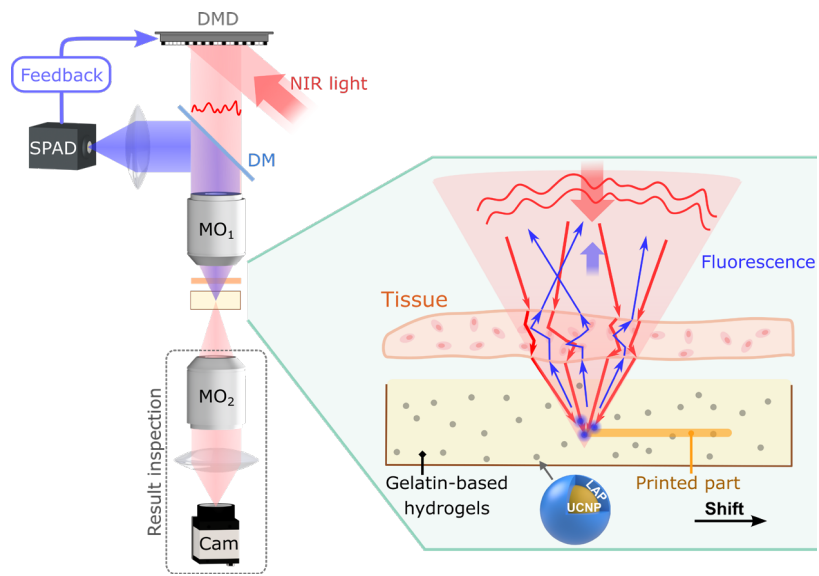

**Fig. 1** Schematic figure of high-resolution non-invasive additive manufacturing using UCNPs that are coated with the photoinitiator LAP. The NIR beam is modulated by the DMD to compensate for the scattering and focus the light through the tissue down to a diffraction-limited spot. The resin contains UCNPs that convert NIR light to UV and visible fluorescence, acting as the secondary UV source for inducing the photopolymerization of the hydrogels and as a guide star for wavefront shaping. By laterally shifting the sample across the print geometry, micro-meter resolution features can be printed through the tissue. DM: dichroic mirror. MO: microscope objective. Cam: camera.

The focusing process using the upconverted fluorescence as feedback is shown in Fig. 2. UCNPs convert NIR light to UV and visible light (Fig. 2a). The latter conveys information about the NIR speckle pattern within the resin. Fig. 2b shows the emission spectrum of UCNPs illuminated by 976-nm light and the absorption spectrum of the photoinitiator LAP. The emission peaks at 350 and 360 nm fall within the absorption band of LAP, suggesting that it is mainly absorbed by the LAP coating and contributes to photopolymerization. The rest of the fluorescence can be partially detected in reflection thanks to its isotropic emission. Upconverted fluorescence in the wavelength range of 440 nm < λ < 550 nm is experimentally chosen as the feedback for the following consideration. The upconversion process to a high-energy photon involves multi-photon absorption, resulting in a nonlinear luminescence process. Each fluorescence peak corresponds to a certain nonlinearity parameter $n$, which can be understood as the number of NIR photons absorbed required to emit a photon of higher energy than the incident NIR photons. Due to the saturation of the excited energy states, the nonlinearity is experimentally experienced only at low light intensity. In the focusing process, signals generated from a high nonlinearity conversion are preferred because of their faster converging speed[37,45,46]. However, this signal only occurs at low intensity and shorter wavelength[47] and thus there is a balance between non-linearity and signal intensity since a higher photon count enables a faster collection and speeds up the focusing process. Upconverted fluorescence in this wavelength range is chosen because it covers most of the photons from visible emission and preserves a high average non-linearity. The total fluorescence is measured by the SPAD at different NIR intensities and plotted in the log scale (Fig. 2c). The NIR intensities are calculated by the laser power measured before the illumination objective divided by the beam size at the focal plane, and the transmission of this objective (~67% at 976 nm) is not taken into account. According to the definition, the slope of the curve in log scale represents the nonlinear parameter $n$. The total fluorescent signal collected displays a slope of 2.4 at a lower intensity and a decreased nonlinearity at a higher intensity.

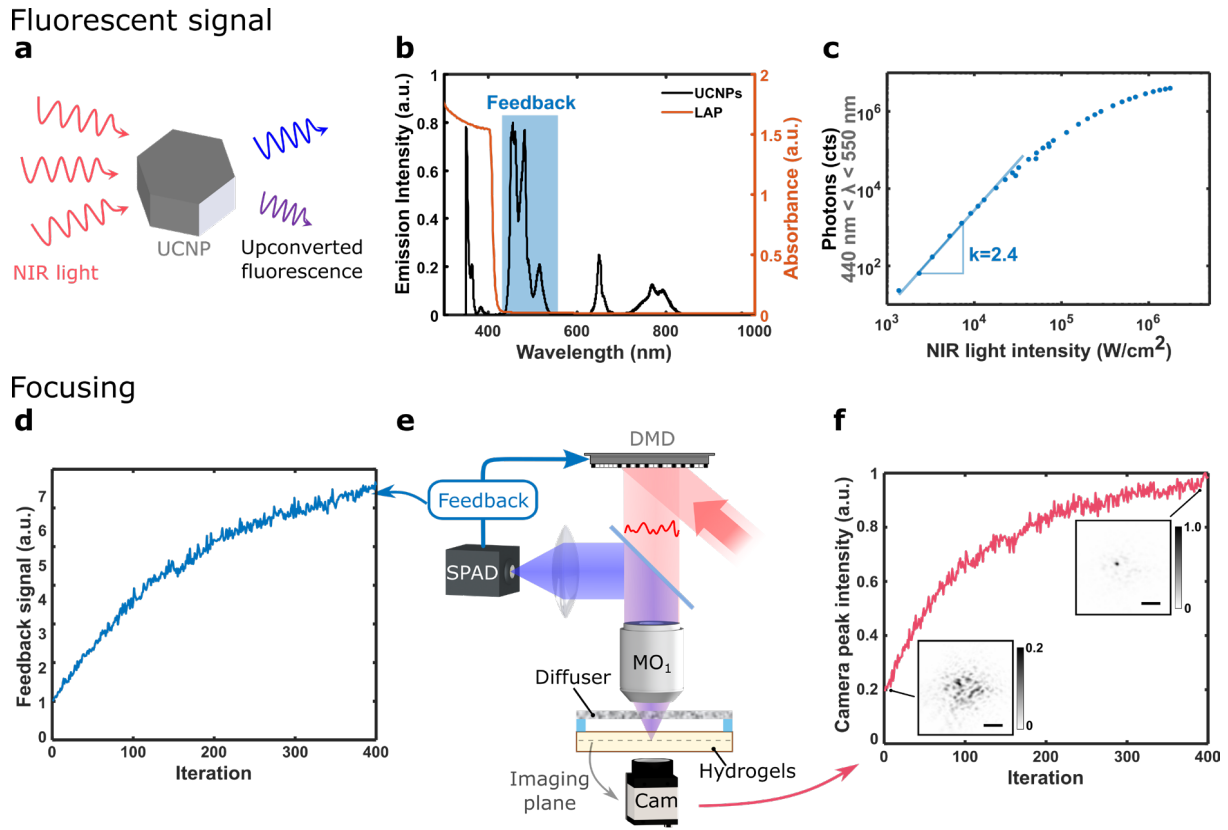

**Fig. 2** Focusing process based on the nonlinear fluorescent feedback. **a** Schematic diagram of the upconverting process. **b** Emission spectrum of UCNPs (black) under 976-nm light illumination and absorption spectrum of the photoinitiator LAP (orange). The wavelength band highlighted in blue represents the range of upconverted fluorescence collected as the feedback. (a.u.= arbitrary units). **c** Fluorescent feedback versus NIR light intensity in log scale. The slope represents the nonlinearity parameter $n$. **d** Power-corrected feedback signal during the iterative optimization. **e** Experimental setup for non-invasive focusing. A holographic diffuser is used in this experiment. **f** Peak intensity of the NIR patterns inspected by the camera during the optimization. Insets show the NIR patterns at the first and the final iteration. Scale bars, 10 μm.

Because of the low upconversion efficiency of UCNPs, the total fluorescence is detected by a single-pixel detector to ensure a good signal-to-noise ratio, at the expense of a loss of spatial information of the speckle. As already demonstrated[37,45,48], optimizing a nonlinear spatially integrated signal enables blind focusing behind a scattering layer through iterative optimization. By maximizing the total fluorescent signal, the light tends to redistribute the energy to one single spot rather than over several grains of a speckle thanks to the nonlinear fluorescence behavior at the chosen NIR intensity. Note that, we have no control over the position of the focal spot, which can be at any hot spot of the speckle illuminating the resin. In most of the previous research works[37,45,46,48], the light is modulated in phase with a liquid crystal-based spatial light modulator (LC-SLM), which does not change the light power after the modulation. In this work, however, a DMD is implemented because of its faster operation (~20 kHz) compared to that of LC-SLM (~60 Hz). Therefore, the light is modulated only in amplitude, resulting in pattern-dependent output power. The fluorescence signal is also dependent on the number of pixels on the DMD with the "ON" state, which does not necessarily result in a focal spot. Therefore, the fitness function $f(x)$ for a DMD pattern $\mathbf{Z}$ is calculated as:

$$f(\mathbf{Z}) = \frac{P_{fluo}}{\mathbf{I}_{DMD} \odot \mathbf{Z}} \quad (1)$$

$P_{fluo}$ is the total fluorescent signal resulting from this DMD pattern. $\mathbf{I}_{DMD} \odot \mathbf{Z}$ is the element-wise product of NIR light distribution on DMD (Fig. S1) and the DMD pattern $\mathbf{Z}$, which gives the light power of this pattern before the illumination objective. By maximizing this fitness function, the iterative algorithm tries to find the DMD pattern that excites more fluorescence per NIR light power, which compensates for the effect of amplitude modulation. In the iterative optimization, we adopted separable natural evolution strategies[49] (SNES) to increase the converging speed of the global search and shorten the optimization time. Multiple pixels are encoded with a number between 0 and 1 (Fig. S2a) to eliminate the drastic change between pixels in the binary amplitude modulation[49]. The focusing process is operated at low NIR power (average intensity of ~3×10³ W/cm²): the nonlinearity parameter $n$ is large, resulting in a faster converging speed; the light dose is much lower than the photopolymerization threshold so that it does not induce photopolymerization. The camera placed on the other side of the sample is only used for imaging the NIR pattern. Because the focal spot can converge at any position of the resin volume illuminated by the speckle, the resin is contained in a rectangular capillary with an inner thickness of 20 μm to limit the position of the focal spot along the optical axis, making it easier for the alignment of the imaging system (see Section S3, Supplementary Information). The peak intensity on the image of the NIR pattern increases with the iteration (Fig. 2f), leading to only one sharp spot behind the diffuser.

After forming a sharp spot behind the diffuser, the optical memory effect can be measured by shifting the sample laterally (Fig. 3a). The capillary containing the resin is fixed onto the diffuser by a spacer (1 mm). Shifting the diffuser together with the resin is equivalent to shifting the beam except that the focal spot will remain at the same position on the image captured by the camera, making the inspection easier. Fig. 3b shows that the peak intensity of the focal spot decreases with the distance $\Delta x$ from the original position. For this holographic diffuser, the full width at half-maximum (FWHM) of the memory effect range is 16 μm.

## Optical memory effect

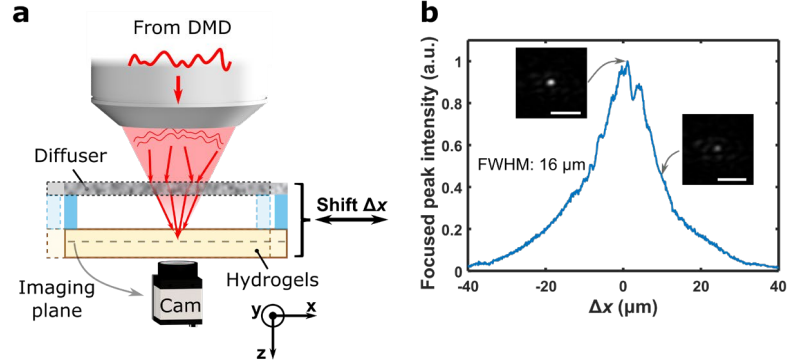

## Dynamic focusing

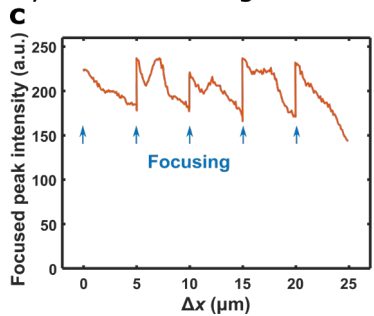
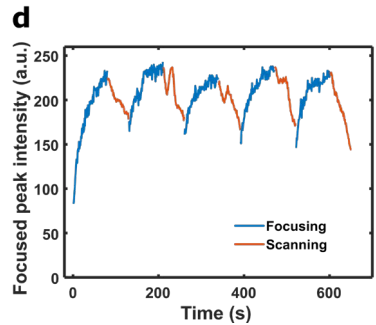
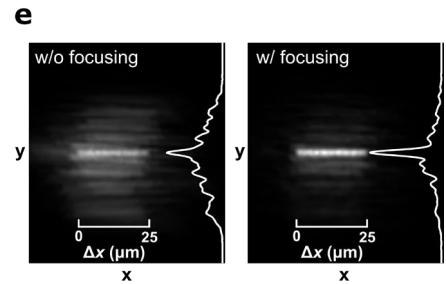

**Fig. 3** Dynamic focusing and scanning based on the optical memory effect. **a** Schematic figure of the setup showing the lateral shifting of the sample. **b** Focused peak intensity with the DMD pattern from the initial focusing versus the shifted distance for the holographic diffuser. Insets show the focal spot at $\Delta x = 0$ and $\Delta x = 10$ μm. Scale bars, 10 μm. Focused peak intensity versus the shifted distance (**c**) and time (**d**) during the dynamic focusing. The scanning process results in a decrease in peak intensity at the focal spot. The focusing process is repeated every 5 μm to maintain a sharp and intense focus. **e** NIR light dose without and with focusing after a shifted distance of 25 μm. The curve on the right-hand side of each image shows the NIR intensity profile along $\Delta x = 12.5$ μm.

Thanks to the memory effect, we are able to scan the focal spot within the field of view without changing the DMD pattern at each printed voxel. However, the spot intensity decreases with the shifting distance, as well as the contrast (Fig. S4), and a high-resolution structure can only be fabricated within a small area around the initial focusing position. In addition, the light dose is not uniform, resulting in different degrees of polymerization across the structure. To maintain a similar light dose at each voxel to be printed, we adopt dynamic focusing. The focused spot is shifted across the diffuser over 5 μm before the re-focusing process starts again (Fig. 3c). The optimized pattern from the previous focusing process serves as the initial pattern for the new optimization, which greatly increases the converging speed compared to the optimization from scratch[44]. The peak intensity of the first iteration of the focusing is lower than that of the ending position of the last scanning process (Fig. 3d) because random deviations are introduced to the initial pattern in order to find the global maximum. Because of the low fluorescent signal collected, the speed of the focusing process is limited to the integrating time of the SPAD for each display. A DMD framerate of 300 Hz is used to ensure that the signal has enough signal-to-noise ratio to reflect the information of the speckle. The focusing time is approximately 90 s (limited by the fluorescence collection) and the scanning time is approximately 50 s (limited by the required dose to solidify the resin) for each 5 μm of lateral shift. The focal spot maintains a relatively stable intensity during the whole process. Fig. 3e shows the NIR light dose distribution on the focal plane with and without dynamic focusing. The experiment without focusing is conducted by laterally shifting the speckle over the same total distance (25 μm). The light dose distribution is calculated by summing up the NIR speckle patterns during the scanning process according to the shifting distance. The profile at $\Delta x = 12.5$ μm is plotted on the right-hand side of each image to show the contrast. With dynamic focusing, we are able to create a dose distribution in the shape of a sharp line with a uniform intensity.

Invasive printing is first conducted in order to explore the possibilities of this technique in tissues without the limitation of the low fluorescent signal after back-scattering. It is demonstrated using both a holographic diffuser and a slice of 300-μm-thick chicken breast. We call this invasive printing because the SPAD detector is placed on the distal side of the sample (Fig. 4a). It is worth stressing that this is the only case in this study, in which the feedback signal is collected in a transmission-based configuration. As the fluorescence is directly collected from the emission site, and thus not experiencing strong loss due to back-scattering, the integration time of the SPAD can be significantly decreased, reducing the optimization time down to 20 s (the DMD displays patterns at 1 kHz during the iterative optimization). The framerate is limited by the long rise time (~0.2 ms) and decay time (~0.3 ms) of the upconverted luminescence of our UCNPs (see Section S5, Supplementary Information). With a faster light modulation, multiple illumination patterns might contribute to the measured fluorescent signal, resulting in inaccurate feedback. The optical memory effect of the chicken tissue has a FWHM of 8 μm, smaller than that of the diffuser (Fig. 4b). Fig. 4c and 4e show qualitatively the impact of scattering on the text readability placed underneath for the holographic diffuser and the chicken breast layer respectively. Letters "EPFL" are printed to verify the capability of printing length scale several times the optical memory lateral shift. During the printing, the focusing process is performed at low NIR power (to benefit from the non-linearity of the fluorescence signal), and the lateral scanning is performed at high NIR power to ensure that the light dose at the focal spot surpasses the polymerization threshold. Fig. 4d and 4f show the printed structures imaged by a differential phase contrast (DPC) microscope[50], which is used in this study because of the low refractive index mismatch between the polymerized and unpolymerized hydrogels (see Section S6, Supplementary Information). The bright and dark edges in the DPC image represent the distribution of phase change (refractive index mismatch) and its contrast (bright minus dark intensity) is positively correlated with the strength of phase change, hence the degree of photopolymerization in this study. The initial focal spot of each letter (top-left corner) is optimized from scratch (a speckle pattern) and the rest is completed with dynamic optimization (from a dim focal spot). The printed structure through the holographic diffuser is relatively uniform, matching the result of the light dose (see the supplementary video). Looking in detail, the line at the bottom of the letter "E" is detached from the rest of the letter at the bottom-left corner. This is because we have no control over the position of the focal spot (global maximum), the position of which happens to switch at this position. For the printed part through the chicken tissue, over-polymerization can be seen in the letter "P" and "F", while optimization did not converge completely when printing the letter "E". It is very likely that the non-uniformity of muscle fibers in the chicken breast results in different scattering properties across the tissue. The tissue structure above the letter "E" is probably more scattering and requires longer focusing time while the tissue above letters "P" and "F" is less scattering, resulting in a brighter focal spot and a higher degree of polymerization. This may be improved by real-time adjustment of the laser power and the optimization parameters according to the fluorescent signal detected.

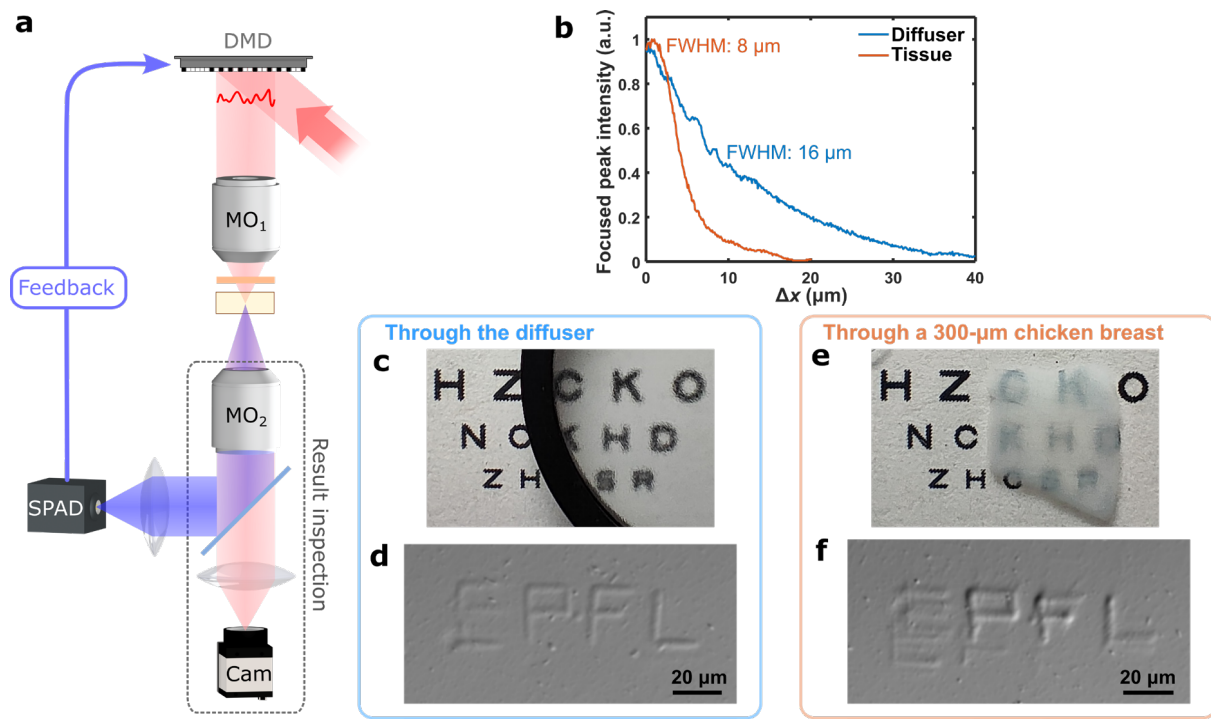

**Fig. 4** Invasive printing. **a** Experimental setup for invasive printing. **b** Focused peak intensity versus shifted distance for the holographic diffuser (blue) and for the chicken tissue (orange). Photographs of a target placed below the diffuser (**c**) and the tissue (**e**). These scattering media prevent visually differentiating the letters of the bottom line. DPC images of printed "EPFL" through the diffuser (**d**) and the tissue (**f**).

Non-invasive printing is then demonstrated through the diffuser. The chicken tissue is not tested in this configuration: the tissue is too much scattering at shorter wavelengths, significantly decreasing the amount of epi-detected (reflection mode) fluorescence which translates into a too-low optimization speed for printing. Based on dynamic focusing, we are able to print fine structures within the speckle (Fig. 5b). Without refocusing (DMD acts only as a mirror), only hot spots in the speckle are printed. By shifting the sample laterally, lines of different contrast and lengths are photopolymerized within the areas highlighted by dash lines (Fig. 5c). As the intensity of the speckle grains decreases at the edge of the speckle, there is no sharp boundary between the polymerized and unpolymerized area, deteriorating the printing fidelity and the printing resolution. The sample is also laterally shifted along the path of a letter "E". As expected, no resolvable structures can be printed inside the speckle size without focusing (Fig. 5d). In contrast, with focusing, sharp and uniform lines with a feature size of ~ 1.5 μm (Fig. S6b) can be printed and the minimum resolvable distance that we obtained is 2 μm (Fig. 5e). Fig. 5f shows a clearly printed letter "E" which is even smaller than the speckle size. Although we have no precise control over the absolute position of the printing (it may start at any place inside the speckle), the relative position of the structures is controlled accurately.

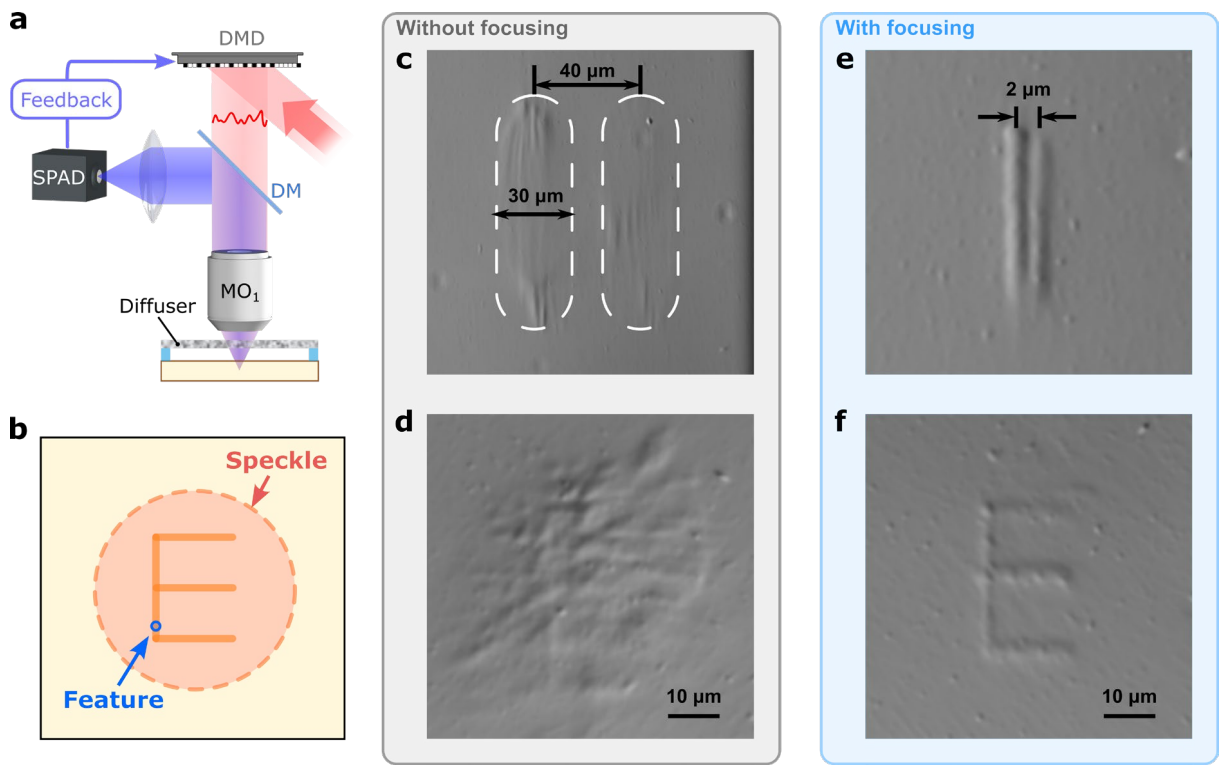

**Fig. 5** Non-invasive printing. **a** Experimental setup for non-invasive printing. **b** Schematic diagram of a comparison of the speckle size and the printed feature size. Fine structures can be printed accurately inside the speckle. **c** DPC image of two "lines" printed without focusing and at a distance of 40 μm. The polymerized area

is roughly highlighted by white dash lines because no clear boundary can be observed. **d** DPC image of a polymerized part following the path of the letter "E" without focusing. **e** DPC image of two printed lines with a center-to-center distance of 2 μm using dynamic focusing. **f** DPC image of a printed letter "E" using dynamic focusing.

## Discussion

Focusing light through strongly scattering media had long been considered impossible until the recent progress in the field of wavefront shaping. Driven by the ever-growing need for deep tissue *in vivo* imaging[29,51], non-invasive imaging has been achieved through various techniques such as blind focusing with nonlinear signals[37,48], fluorescence-based transmission matrix[40,52], and acoustic manipulation[39]. In this work, we use blind focusing which utilizes nonlinear signals to cope with the issue of low feedback signal. This has been demonstrated with two-photon imaging[37] and three-photon imaging[45,48] and should be also available for other nonlinear processes[53]. For linear signals that have enough signal intensities, non-invasive focusing and imaging can be achieved by collecting the back-scattered fluorescent patterns[40,52]. The techniques readily available in the wavefront shaping might inspire the development of new bioprinting methods against the turbid nature of the biological tissue.

In this work, we make use of an iterative wavefront method to enable non-invasive additive manufacturing through a scattering layer. The light propagating through the scattering medium produces complicated speckle patterns that locally excite UCNPs. Because of the nonlinear upconverted fluorescence as the feedback, a sharp focus can be formed in the resin at a fast optimization speed even with a low signal level. Dynamic focusing is conducted to ensure a uniform light dose to solidify voxels and combined with the memory effect, the printing time is optimized. A printing scale 5 times the memory effect size is demonstrated. Successful printing through a diffuser and a 300-μm chicken tissue proves the feasibility of this technique. In the non-invasive configuration, we show that a micro-meter resolution structure can be printed within a speckle pattern that is more than 20 times larger than the feature size. To the best of our knowledge, this study is the first report on high-resolution non-invasive printing through a highly scattering medium, which pushes the boundaries of noninvasive printing to micro-meter resolution.

There is much work to be done before the proposed technique can become a tool for non-invasive *in vivo* printing. Biological tissues are dynamic scattering media[39,54,55] and therefore the time required to focus light in our approach cannot exceed the ms range. The optimization speed in the non-invasive focusing is limited by the signal intensity, which can be increased by using a larger and more sensitive photon detector such as photomultiplier tubes and more importantly improving the upconversion efficiency of the nanoparticles[56]. SPAD arrays might also be used to gather the spatial distribution of the fluorescence. Noise-tolerant algorithms with relatively fast converging speed are preferred and deep learning might also be implemented for efficient focusing through living tissues[57].

Real applications also pose requirements in the spatial domain. The nature of blind focusing in this technique denies the possibility of pre-determining the absolute position of the voxel. In this work, we observe that the global maximum will remain at a hot spot for approximately 25 μm of shifting in the lateral direction before switching to another hot spot within the illuminated resin. This determines the printing area of one object (see Section S7, Supplementary Information). The optical memory effect also exists in the axial direction[58], indicating the possibility of true 3D printing. As for a larger volume, recent progress in imaging beyond the memory effect[52] might help to push this boundary. For applications that require an accurate absolute position of the printed voxels such as connecting the neuron fiber, feedback with spatial information is necessary.

In summary, we have presented a non-invasive additive manufacturing technique to print a hundred of micron structure size through a strongly scattering medium at micro-meter resolution based on the fluorescent feedback from the printing system. Thanks to the nonlinear upconverted fluorescence and the optical memory effect, sub-speckle printing is demonstrated on a 25-μm size print with a printing resolution of around 2 μm. This technique provides a promising route toward high-resolution non-invasive bioprinting and shines light on the development of new techniques for minimally invasive and non-invasive biomedicine.

## Materials and methods

**Synthesis of NaYF$_4$:Yb/Tm core UCNPs and NaYF$_4$ shell precursor**

Chemicals used in this experiment were purchased from Merck & Co (Sigma-Aldrich) and the synthesis was carried out in a bifold Schlenk line under the flow of argon gas.

In a typical synthesis, thulium (III) acetate hydrate (0.004 mmol) was reacted with oleic acid (6 mL) and 1-octadecene (15 mL) at 140 °C under partial vacuum having argon atmosphere for 90 min in a 100 mL 3-neck Schlenk flask to prepare oleate solution. Once the reaction was complete, first, ytterbium (III) acetate hydrate (0.240 mmol) was reacted with the above oleate solution for 90 min and afterward, yttrium (III) acetate hydrate (0.556 mmol) was reacted to this oleate solution at 140 °C. This mixed oleate solution, thus obtained, was cooled down to 50 °C. To this, methanol solution (10 ml) of ammonium fluoride (3.2 mmol) and sodium hydroxide (2 mmol) was added dropwise and stirred for 30 min. Methanol was completely removed under partial vacuum and the reaction mixture was further heated to 300 °C (~10 °C/min) under argon and maintained for 60 min. The reaction was frozen by the addition of cold ethanol and the NCs were collected by centrifugation, redispersed in cyclohexane. This process was repeated thrice, before being used as core UCNP (0.5 mol% $Tm^{3+}$, 30 mol% $Yb^{3+}$ doped) in the next step.

Similarly, in the second step, yttrium (III) acetate hydrate (0.8 mmol), oleic acid (6 ml) and 1-octadecene (15 ml), methanol solution (10 ml) of ammonium fluoride (4 mmol) and sodium hydroxide (2.5 mmol) were used to prepare the $NaYF_4$ shell precursor.

**Synthesis of ligand free $NaYF_4$:Yb/Tm @ $NaYF_4$ core-shell UCNPs**

Layer-by-layer successive epitaxial shell growth of $NaYF_4$ was achieved on $NaYF_4$:Yb/Tm core UCNPs. Core UCNPs were added to 1-octadecene (5 mL) in a 3-neck Schlenk flask and heated to 300 °C in an argon atmosphere. To this, shell precursor solution was injected @ 5 µL/sec using a nemesys syringe pump system. The ripening was done at 300 °C for 30 min. After ripening, the reaction was frozen and the core-shell UCNCs were precipitated and washed as outlined for core UCNPs and finally dispersed in hexane (5 mL). These dispersed particles were neutralized using 2M HCl to get the ligand-free core-shell UCNPs.

**Synthesis of gelMA**

10 g of gelatin (Sigma-Aldrich) was dissolved in 100 mL of phosphate-buffered saline. Then 8 mL of methacrylic anhydride (Sigma-Aldrich) was added dropwise (0.5 mL/min) and the mixture was left under stirring at 50 °C for 3 hours, followed by removal of unreacted anhydride by centrifugation and dialysis against distilled water. GelMA was obtained after lyophilization.

**Preparation of UCNP-loaded hydrogel**

Lithium phenyl-2,4,6-trimethylbenzoyl-phosphinate (LzAP) (Sigma-Aldrich) was dissolved in water at a concentration of 20 mg/mL. 10 µL of UCNP aqueous solution (100 mg/mL) was mixed with 50 µL of LAP solution and sonicated for 30 min. 15 mg of gelMA was dissolved in the mixture and 40 μL water was added to form a final concentration of 10 mg/mL UCNP, 10 mg/mL LAP and 15 wt% gelMA. The resin was stored at 4 °C until further use.

**Characterization**

The UV-Vis spectrum of LAP was recorded on a Lambda 365 UV/Vis spectrophotometer. The upconverted fluorescence emission spectrum was recorded on a setup as previously reported[47]. DPC images were recorded on a microscope as previously reported[50].

**Experimental setup**

A continuous-wave laser at 976 nm (900 mW, BL976-PAG900, Thorlabs) with a Polarization-Maintaining (PM) optical fiber is collimated by a lens (F810APC-1064, Thorlabs). After modulated by the DMD (V-650L, Vialux), the NIR light is directed through the objective $MO_1$ (M Plan Apo NIR 20X, NA 0.40, Mitutoyo) to excite the UCNPs in the resin placed below the scattering medium. The DMD is imaged to the back focal plane of $MO_1$. The scattering medium is a holographic diffuser (Newport 5°) or a slice of fixed chicken breast. The upconverted fluorescence is back-scattered by the medium, collected by $MO_1$ and a lens (f = 30 mm), and detected by a SPAD (PDM-50-CTD, Micro Photon Devices). We use a dichroic mirror longpass 550 nm and two shortpass filters: a 600-nm shortpass and a 700-nm shortpass to narrow the spectral bandwidth. The NIR speckle patterns are imaged in transmission via $MO_2$ (LIO-40X, NA 0.65, Newport) and a lens (f = 150 mm) onto the Cam (acA2040-55um, Basler). This part of the setup is only used for monitoring the NIR speckles. In the invasive configuration, the fluorescence is collected in transmission through $MO_2$ and a lens (f = 30 mm) and detected by the SPAD.

**Focusing**

The optimization was done with the SNES algorithm[49]. The segment number is 32×17 and the segment size is 25×50, which means that a range of 800×850 pixels on the DMD is used for light modulation. The light distribution on DMD was calibrated by sequentially turning on each segment and measuring the difference in the output power. In each segment, the number of encoded pixels is 5, and the coding strategy is as previously reported[49]. It is encoded in the $x$-axis and expanded to the size of a segment by repeating each pixel in $x$- and $y$-axis (see Section S2, Supplementary information). The population size is 40 and the iteration is 200 for invasive focusing and 400 for non-invasive focusing. These parameters are chosen to balance the optimization speed and the enhancement. The DMD display speed is mainly limited by the fluorescent signal intensity. 1 kHz is used in the invasive configuration and 200~300 Hz is used in the non-invasive configuration.

**Printing**

The UCNP-loaded hydrogel was sonicated at 40 °C for 1 min before it was filled into the rectangular capillary (20 μm × 200 μm). The capillary was fixed onto the diffuser, which was pasted on a glass slide in order to be clamped by the sample holder. The distance between the holographic diffuser and the resin is 1 mm in air and 14 μm of capillary glass wall; the distance between the chicken tissue and the resin is 170 μm of the coverslip and 14 μm of the glass wall. A "white" pattern was displayed on the DMD, making it just as a mirror. The sample was aligned in the *x*, *y*, and *z* direction so that the capillary is illuminated by the speckle and the speckle size within the capillary is about 30~50 μm. Then the SPAD was aligned to maximize the fluorescent signal.

During the printing, the focusing process was conducted at low power (~7 mW before $MO_1$), and then the focal spot was scanned at higher power (~25 mW before $MO_2$) in order to surpass the polymerization threshold. The hatching distance is 1 μm and the scanning speed is 0.1 μm/s.

**Tissue fixation**

A piece of fresh chicken breast was cut into 4-mm cubes and fixed with 10% buffered formalin (HT501128-4L, Sigma Aldrich) overnight. Then the fixed samples were rinsed in phosphate-buffered saline three times and embedded in 2% agarose until solidified. The embedded tissues were cut into 300-μm-thick slices using a vibratome (VT1200 S, Leica), and mounted onto glass slides (Sigma-Aldrich) with Fluoromount-G (SouthernBiotech). 300-μm spacers were used to confine specimens without compression. The sections were sealed with nail polish and kept at 4 °C for 24 hours before being used for printing.

# Acknowledgment

This project has received funding from the Swiss National Science Foundation under project number 196971 - "Light based Volumetric printing in scattering resins." The authors thank the open-source tools (and their contributors) that were used in this work, including Inkscape.org, Python.org, and Fiji.sc. The authors would like to acknowledge Dr. Jessica Sordet-Dessimoz (Histology Core Facility, EPFL) for assistance with tissue sectioning. The authors also acknowledge Dr. Jorge Madrid-Wolff for discussions and constructive comments.

# Conflict of interest

The authors declare no conflicts of interest regarding this article.

# Supplementary Information: Light focusing and additive manufacturing through highly scattering media using upconversion nanoparticles

**Qianyi Zhang[1], Antoine Boniface[1], Virendra K. Parashar[2], Martin A. M. Gijs[2], and Christophe Moser[1, *]**

[1] Laboratory of Applied Photonics Devices, School of Engineering, Institute of Electrical and Micro Engineering, Ecole Polytechnique Fédérale de Lausanne, Lausanne, Switzerland

[2] Laboratory of Microsystems LMIS2, School of Engineering, Institute of Electrical and Micro Engineering, Ecole Polytechnique Fédérale de Lausanne, Lausanne, Switzerland

[*]To whom correspondence should be addressed. e-mail: christophe.moser@epfl.ch

## S1: NIR light distribution on DMD

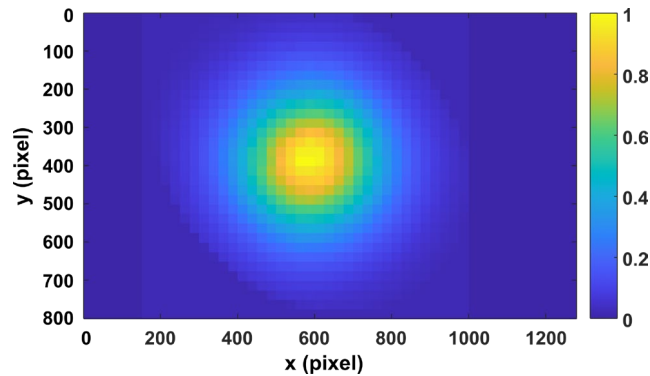

**Fig. S1**. Normalized NIR light distribution on DMD. It is calibrated by flipping each segment (25 × 25 pixels) of DMD and measuring the output power change.

## S2: DMD encoding strategy

In the iterative algorithm, the search patterns are grayscale patterns (between 0 and 1) with a matrix size of 32 × 17. Then each element of the pattern is encoded in a rectangular segment (25 × 50 pixels) on DMD. The encoding strategy is shown in Fig. S2. First, this element is encoded with 5 binary pixels using a lookup table illustrated in Fig. S2a. Then these 5 pixels are expanded to a segment by repeating each pixel for 10 times in $y$-axis and 25 times in the $x$- axis. For example, an element 0.25 will be encoded by a segment pattern shown in Fig. S2b. Fig. S2c shows a DMD binary pattern encoding a grayscale search pattern with this strategy.

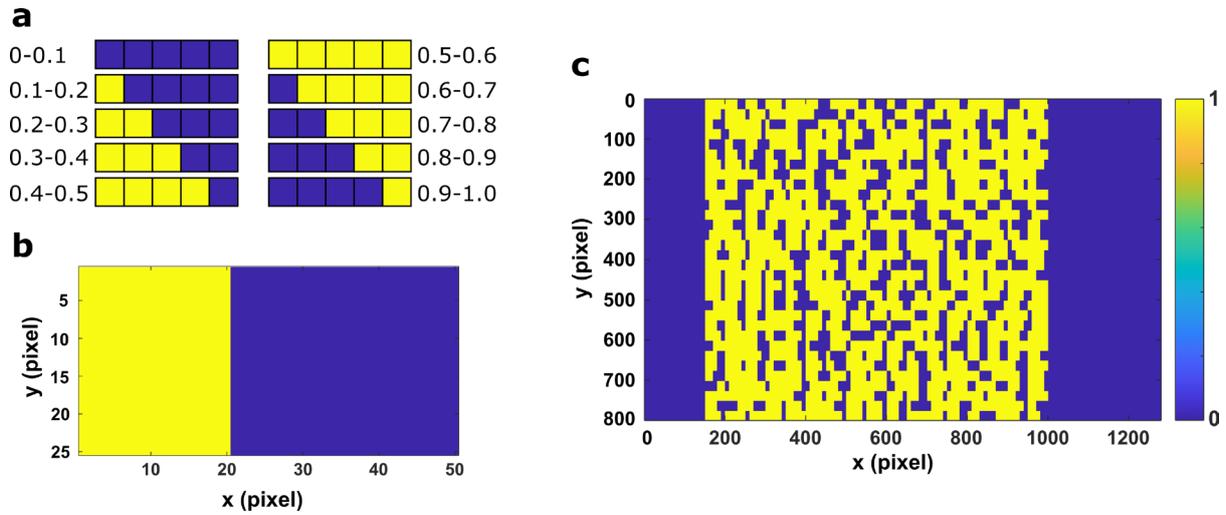

**Fig. S2**. **a** Lookup table of the multi-pixel encoding strategy. **b** A segment that can encode a number between 0.2 and 0.3. **c** An example of a DMD binary pattern generated using this encoding strategy.

## S3: The position of the focal spot in axis direction

In this experiment, a square capillary (100 μm × 100 μm) filled with the resin is used instead of the rectangular capillary. The SPAD is placed on the distal side of the diffuser (invasive focusing) for a faster optimization speed. After a focal spot is formed by the optimization, the objective $MO_2$ is aligned to image the focal plane of this spot. Then $MO_2$ is moved along the optical axis (*z*-axis) by a motorized stage and the transverse intensity profile at each *xy*-plane is recorded by a camera. Fig. S3a shows the longitudinal section of the focused beam profile. The z-position of this focal spot is set to 0 for comparison. After the measurement, $MO_2$ is moved back to the initial position (the focal plane of the initial spot). The sample is shifted in *y*-direction for 100 μm (beyond the optical memory effect) to position 2 to ensure that the focal spot will switch to another hot spot. After an optimization from scratch, the beam profile is measured again. Fig. S3b shows that this focal spot appears at the plane z = 25 μm. This means that $MO_2$ needs to be adjusted in *z*-direction for a relatively large range to find the focal plane of the spot. An easy solution is to use a 20-μm-thick resin confined by the rectangular capillary (20 μm × 200 μm). The axial feature size of the focal spot is 10~15 μm. Therefore, the focal spot is around the center of the capillary, making it easier to find the focal plane for imaging.

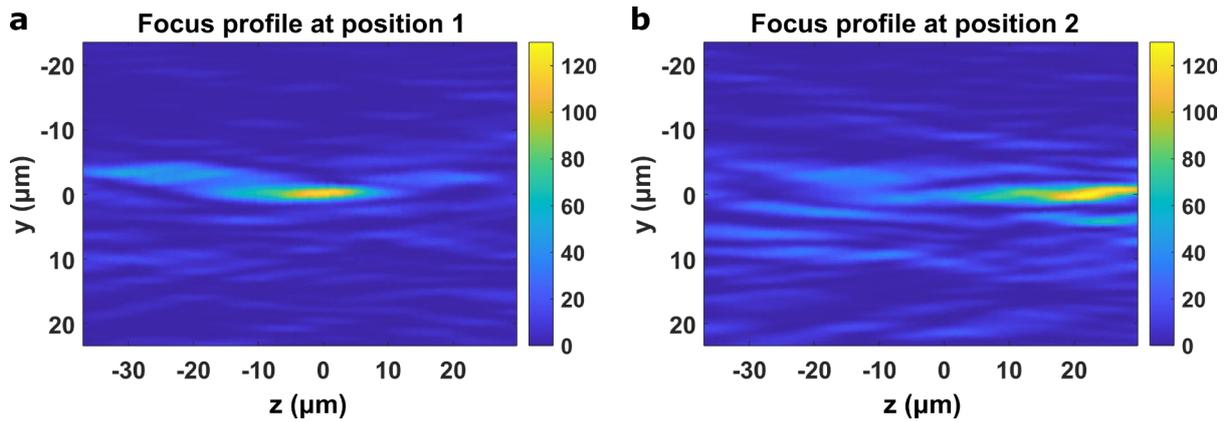

**Fig. S3 a** NIR beam profile in *yz*-plane at position 1. The focus position in *z*-direction is set as z = 0. Then the sample is shifted in the lateral plane to position 2. **b** NIR beam profile of the focusing result at position 2. Absolute values are used in the *z*-axis tick for comparison and relative values are used in the *y*-axis tick. It shows that the position of the focal spot can appear in a large range (at least ± 25 μm) in *z*-direction.

This practice does not affect the focusing and printing because the imaging system is only used for inspecting NIR patterns. However, a thicker resin results in more hot spots in the illuminated volume, which will slow the converging speed[1].

## S4: Light dose distribution of sample shifting after one focusing process

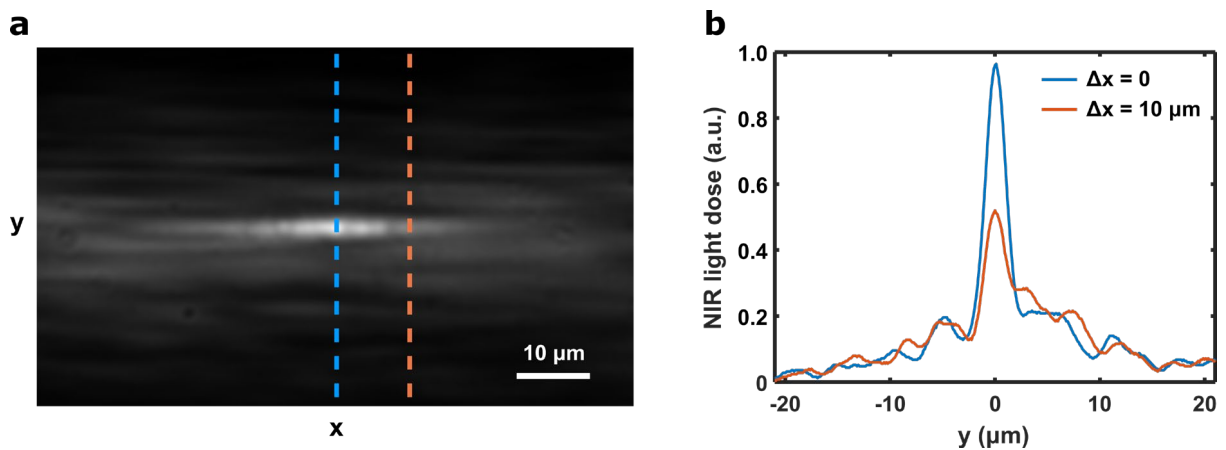

**Fig. S4 a** NIR light dose distribution after one focusing process followed by sample shifting in the *x*-direction. **b** Profile of the light dose along the dash lines marked in (**a**) at $\Delta x = 0$ (blue) and $\Delta x = 10$ μm (orange). The peak intensity as well as the contrast decreases significantly with the shifted distance.

## S5: Responding speed of the fluorescent signal

The rise and decay dynamics are studied using 976-nm excitation pulses. We only studied the dynamics of the emission peak at 475nm due to the availability of the bandpass filter in our lab.

In Fig. S5a, the excitation pulse starts at t = 0 and lasts 12.5 ms. The rise time is defined as the time between the initial rising moment and the moment reaching the plateau of the fluorescent emission. Effective decay time is defined by[2]:

$$\tau = \frac{\int_{t=initial\ decay\ moment}^{\infty}(t - initial\ decay\ moment)I(t)dt}{\int_{t=initial\ decay\ moment}^{\infty}I(t)dt} \quad (1)$$

At an excitation intensity of 10 MW/cm$^2$, the rise time is 0.18 ms and the decay time is 0.32 ms. At the intensity of 1.5 MW/cm$^2$, the rise time is 0.26 ms and the decay time is 0.33 ms. Multiple DMD patterns might contribute to the measured fluorescent signal when operating at a high display rate (1 kHz or higher).

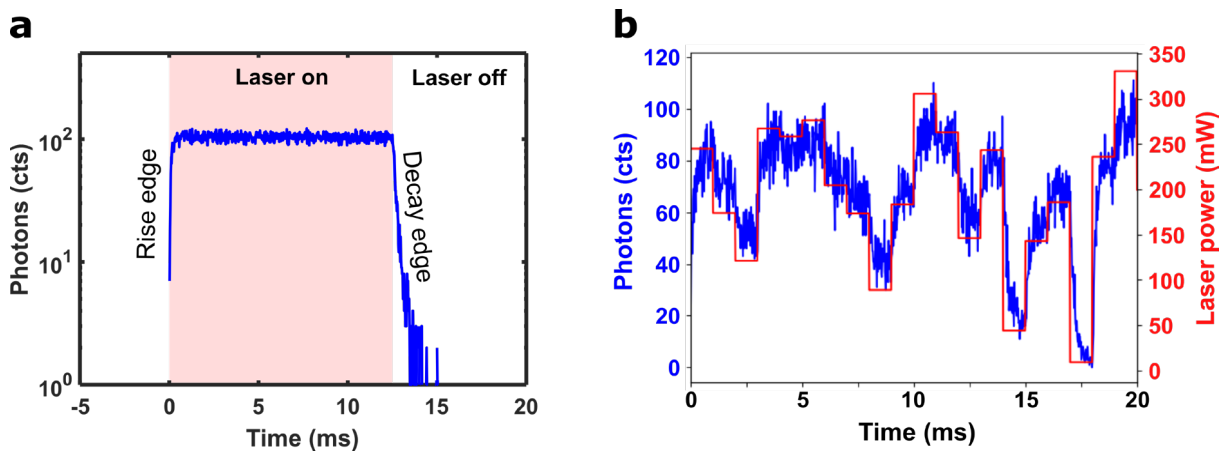

**Fig. S5 a** Rise and decay dynamics of the fluorescent peak at 475 nm. **b** Fluorescent signal response (475 nm) of random laser powers operated at 1 kHz.

We test the responding dynamics of the fluorescent signal using random laser powers modulated at 1 kHz. As the fluorescent intensity increases with the increase of laser power, by comparing the fluorescent signal curve with the laser power curve (Fig. S5b), we are able to check if the fluorescent output can reflect the change of input power. Although it takes time to reach a stable emission, the total fluorescence during the interval of two successive modulations does represent the change of the input power qualitatively. At higher frequencies (for example, 2 kHz), where the time interval (0.5 ms) is similar to the rise and decay time, the fluorescent signal will not be able to reflect the information of the speckle accurately.

## S6: DPC microscopy

Because of the low refractive index mismatch (~0.001) between the polymerized and unpolymerized gelMA, all the printed parts in this study are imaged by a DPC microscope. It consists of a left-half and a right-half LED source. Fig. S6a shows the images taken with these two light sources separately. In the image taken with the left half source, bright edges can be seen on the left side of the printed structure and dark edges on the right side. Then sum of these two images is calculated to mimic the brightfield image and the difference ($I_L$-$I_R$) is calculated to obtain the DPC image. The printed object can hardly be seen in the brightfield image but is visible in the DPC image because the phase change is translated into intensity variation by the asymmetric light source. With light sources placed at the left and the right, the DPC image shows the best contrast in the horizontal direction. Therefore, the sample is placed at 45 ° to obtain a similar contrast across the whole structure. With the knowledge of the light source[3], the quantitative phase image is reconstructed from the DPC image and rotated by -45° for visualization (Fig. S6b). The effect of the optimization process on the obtained print can be seen from dark spots across the printed structures. Although optimized at low NIR power, the long optimization time during noninvasive printing contributes to the photopolymerization light dose, resulting in a larger phase change and, thus higher degree of polymerization. Print feature size is extracted and averaged from the FWHM of the phase profile.

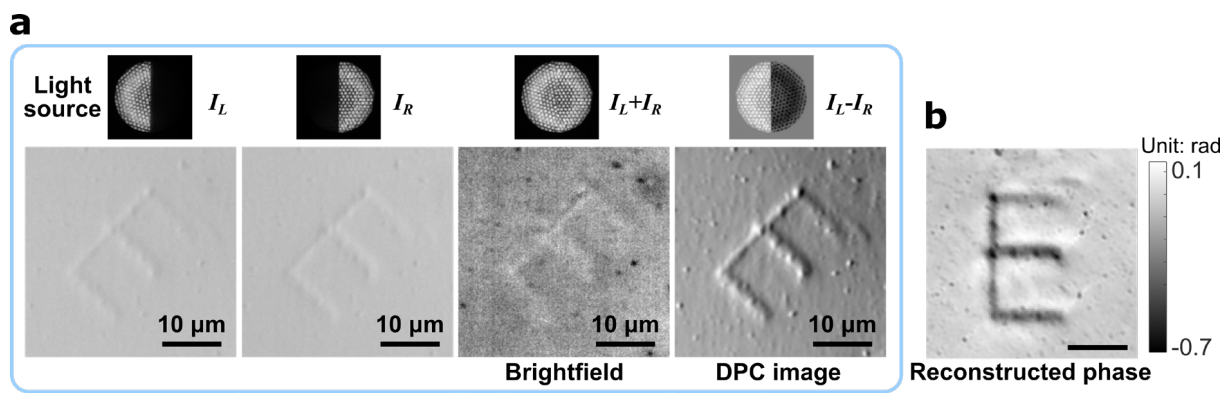

**Fig. S6 a** Intensity image taken with the left and right halves of the LED source. The brightfield image is generated by the sum of the two half-circle images and contains no phase contrast. The DPC image is the difference of the two half-circle images. **b** Reconstructed phase of the DPC image in (**a**) after rotating -45°. Scale bar: 10 μm.

## S7: The position switch of the focal spot

In invasive printing, the fluorescent signal is collected by the SPAD without scattering. The limited active area of the SPAD (d = 50 μm) results in a cylindrical collection volume of ~8 μm in diameter and 20 μm in height (resin thickness). Therefore, the focal spot only switches among the hot spots in this volume (Fig. S7a). In non-invasive printing, SPAD receives fluorescent signals from the whole speckle because of the scattering. The focal spot may switch to a position that is tens of microns away (mainly in the lateral direction) from the initial focal spot during dynamic focusing (Fig. S7b). Fig. S7c shows an example of the focal spot position switch during noninvasive printing. A vertical line was interrupted in the middle by the position switch and a second line was printed at a new position.

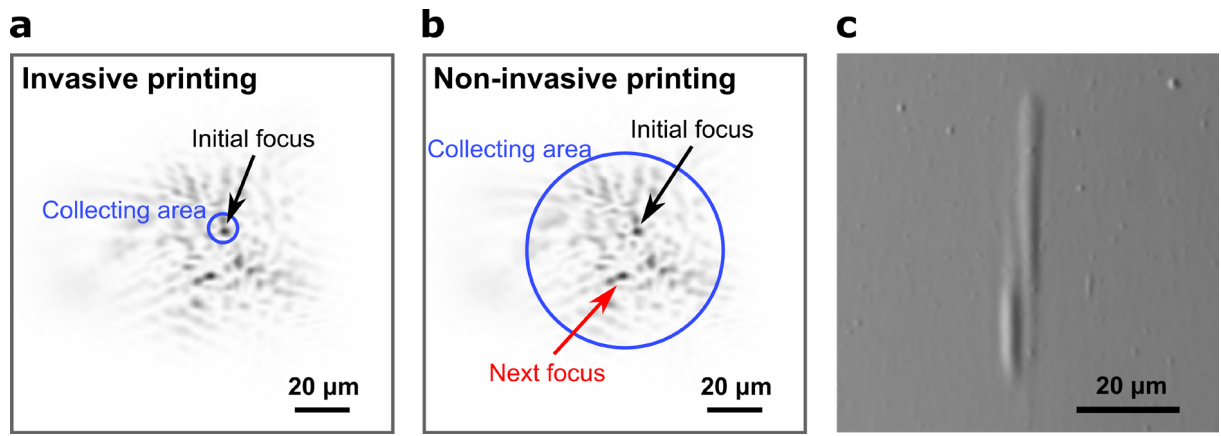

**Fig. S7** Schematic diagram of the focal spot switching during invasive printing (**a**) and non-invasive printing (**b**). **c** A printed structure that was intended to be a 50-μm-long vertical line. A position switch of the focal spot can be observed during the printing.

## S8: Video of the NIR dose distribution during invasive printing of "EPFL"

The video is generated by summing up the NIR patterns based on the shifted position. It is played at 125× of the original printing speed. The focusing process is ignored because it is done at much lower laser power. Fig. S8 shows the last frame of this video.

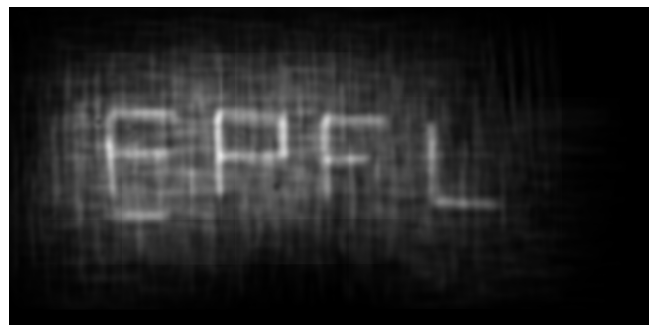

**Fig. S8** Last frame of the video. It shows the final NIR dose distribution of the printing.